# Tensile Properties of Structural I Clathrate Hydrates: Role of Guest-Host Hydrogen Bonding Ability

Yue Xin, Qiao Shi, Ke Xu, Zhisen Zhang[*], Jianyang Wu[*]

Department of Physics, Research Institute for Biomimetics and Soft Matter, Fujian Provincial Key Laboratory for Soft Functional Materials Research, Xiamen University, Xiamen 361005, PR China

**Abstract:** Clathrate hydrates (CHs) are one of the most promising molecular structures in applications of gas capture and storage, and gas separations. Fundamental knowledge of mechanical characteristics of CHs is of crucial importance for assessing gas storage and separations at cold conditions, as well as understanding their stability and formation mechanisms. Here, the tensile mechanical properties of structural I CHs encapsulating a variety of guest species ($CH_4$, $NH_3$, $H_2S$, $CH_2O$, $CH_3OH$, and $CH_3SH$) that have different abilities to form hydrogen (H-) bonds with water molecule are explored by classical molecular dynamics (MD) simulations. All investigated CHs are structurally stable clathrate structures. Basic mechanical properties of CHs including tensile limit and Young's modulus are dominated by the H-bonding ability of host-guest molecules and the guest molecular polarity. CHs containing small $CH_4$, $CH_2O$ and $H_2S$ guest molecules that possess weak H-bonding ability are mechanically robust clathrate structures and mechanically destabilized via brittle failure on the (1 0 1) plane. However, those entrapping $CH_3SH$, $CH_3OH$, and $NH_3$ that have strong H-bonding ability are mechanically weak molecular structures and mechanically destabilized through ductile failure as a result of gradual global dissociation of clathrate cages.

1. Introduction

---

[*]Corresponding Emails: zhangzs@xmu.edu.cn, jianyang@xmu.edu.cn

Clathrate hydrates (CHs) are a non-stoichiometric solid crystalline substance, in which host water molecules are connected through hydrogen (H)-bonding and form cages that accommodate a large variety of small guest molecules[1, 2]. Commonly, CHs form in environments of cold temperature and moderate pressure where water and small molecules coexist, for example, they are abundantly identified in deep ocean floor sediment and on submarine continental slopes, as well as natural gas pipelines[3]. Remarkably, CHs have attracted great scientific and engineering interests as a result of their involvement in potential natural gas resources[4], geological hazards[5], global climate change[6], flow assurance and safety issues in oil and gas pipelines[7], as well as applications of gas storage and separation[8, 9]. For example, it was demonstrated that methane can be rapidly stored in saline water (1.1 mol% NaCl solution) and seawater via clathrate hydrates aided by 5.56 mol% tetrahydrofuran (THF) in a simple unstirred tank reactor[10]. Based on semi-clathrate hydrate (SCH) adsorption media, it was revealed that a two-stage batch separation process allowed enrichment of $CO_2$ to more than 90 mol% from a 20 mol% $CO_2$ feed; however, a three-stage continuous counter current separation process allowed a $CO_2$ recovery of about 90%[11].

Beyond the external conditions of low-temperature and moderate pressure, non-bonded interactions between small guest and water host molecules play the key role on the formation and stability of CHs. It was shown that many factors such as molecular size, atomic charge and polarity of guest molecules, have significant effect on the stability and properties of CHs[12, 13]. As a result of finite three-dimensional (3D) clathrate cages, guest molecules with suitable size trapped in clathrate cages are able to effectively interact with the surrounding water molecules, thereby stabilizing H-bonded water-body skeleton of CHs[14]. Kvamme et al.[15, 16] revealed via Monte Carlo (MC) simulations that structural properties of CHs are influenced by atomic charges of $H_2S$, $SO_2$, and $CO_2$ guest molecules. It was found that, compared with the case of non-charged guest molecules, guest molecules with either average positive or negative charge exposed on the cavity wall enlarge the Langmuir constants. Such effect for $H_2S$ is more pronounced than for both $CO_2$ and $SO_2$. At low temperature condition, CHs entrapping oxygen-contained guest molecules are less structurally stable than those containing hydrocarbons or sulfur[17-19]. Liu et al[20] conducted molecular dynamics (MD) simulations to study the thermodynamic and kinetic feasibility of replacing $CH_4$ from CHs with $N_2/CO_2$ mixtures, and it was found that the replacement of $CH_4$ with $N_2$ and $CO_2$ has a negative Gibbs free energy, however, $CO_2$ molecular substitution in small cages is structurally disadvantageous. In the substitution process, $CO_2$ molecular substitution is dominant in the thermodynamics, while $N_2$ molecular substitution is dominant in the kinetics. Remarkably, the ability of guest molecules to form H-bonds with host water molecules also dictates the structural stability of CHs[21, 22]. For example, it is uncovered via MD simulations that $NH_3$-contained CH show lower decomposition temperature than that of $CH_3OH$-contained one. This is primarily attributed to the fact that $CH_3OH$ is able to form both proton-donating and proton-accepting H-bonds with water, while $NH_3$ mainly forms proton-accepting H-bonds with water.

A number of works have been performed both experimentally and numerically in understanding the stability limits of CHs containing different guest molecules from a mechanical point of view[23-27]. Upon compression tests, it was observed that $CH_4$ hydrate shows different mechanical characteristics from those of water ice, for example, pronounced strain-hardening behaviors accompanied by solid-state disproportionation or precipitation[23, 24]. Subjected to triaxial tension, it was found that ultimate tensile strength of CHs decreases with increasing guest molecular size, and large clathrate cages of CHs are more sensitive to mechanical stress than small clathrate cages[25]. Using MD simulations, $CO_2$ and $CH_4$ CHs show difference in both shear and Young' moduli [26]. Furthermore, through tension MD calculations, it is revealed that the guest molecular size, shape and polarity have significant effects on the mechanical characteristics of CHs[27].

To date, available works concentrating on the effects of guest molecular species on the mechanical properties of CHs are limited. To the best of our knowledge, moreover, the effects of guest-host H-bonding ability on the structural stability of CHs subjected to mechanical loads remain unexplored. Fundamental knowledge of mechanical properties of CHs is of vital importance not only for evaluating the application of gas storage and separations, but also for understanding the stability and formation mechanisms. In this work, tensile mechanical characteristics of CHs entrapping a variety of multielement guest molecular species ($CH_4$, $NH_3$, $H_2S$, $CH_2O$, $CH_3OH$, and $CH_3SH$) that show different abilities to form H-bonds with water are comprehensively investigated by classic MD simulations with TIP4P/ICE and OPLS-AA forcefields.

## 2 Models and Methodology

### 2.1 Molecular Models of CHs

All investigated CHs are water framework of structural I clathrate hydrate. Initial positions of water oxygen atoms of sI clathrate hydrate crystalline lattice are taken from X-ray diffraction data[28]. Based on the Bernal-Fowler rule, hydrogen atoms of water are uniquely assigned with coordinated orientation in the framework, and the dipole moment and potential energy of the H-bond network are minimized[29]. As is seen in Figure 1a, one-unit cell of sI CH is composed of 46 water molecules that forms two $5^{12}$ and six $5^{12}6^2$ polyhedral water cages. In this work, all simulation models are a cubic supercell of 3×3×3 replicas of the sI CH unit cell. All the water polyhedral cages are fully occupied by guest molecules. Periodic boundary conditions (PBC) are employed in all three orthogonal directions to eliminate edge effects and thermodynamic limits. To reveal the effect of guest molecules on the mechanical properties of sI CH, a variety of guest species ($CH_4$, $NH_3$, $H_2S$, $CH_2O$, $CH_3OH$ and $CH_3SH$) that show different dimensionality, molecular polarity and ability of forming H-bonds with water molecule, are selected. Table 1 lists the effective kinetic diameters of those selected guest molecules. Apparently, their effective kinetic diameters are less than 5.0 Å, and thus from the dimensionality alone, both $5^{12}$ and $5^{12}6^2$ water cages are capable of accommodating the investigated guest molecules. Those guest molecules are placed in the center of clathrate cages with random molecular orientation to generate the initial structural models of CHs.

Table 1 Effective kinetic diameters of guest molecules[30-39]

| Guest species | $CH_4$ | $NH_3$ | $H_2S$ | $CH_2O$ | $CH_3OH$ | $CH_3SH$ |
|---|---|---|---|---|---|---|
| Effective kinetic diameter (Å) | 3.8[30-32] | 2.9[33, 34] | 3.6[35, 36] | 3.7[39] | 3.7[35] | 4.0[37, 38] |

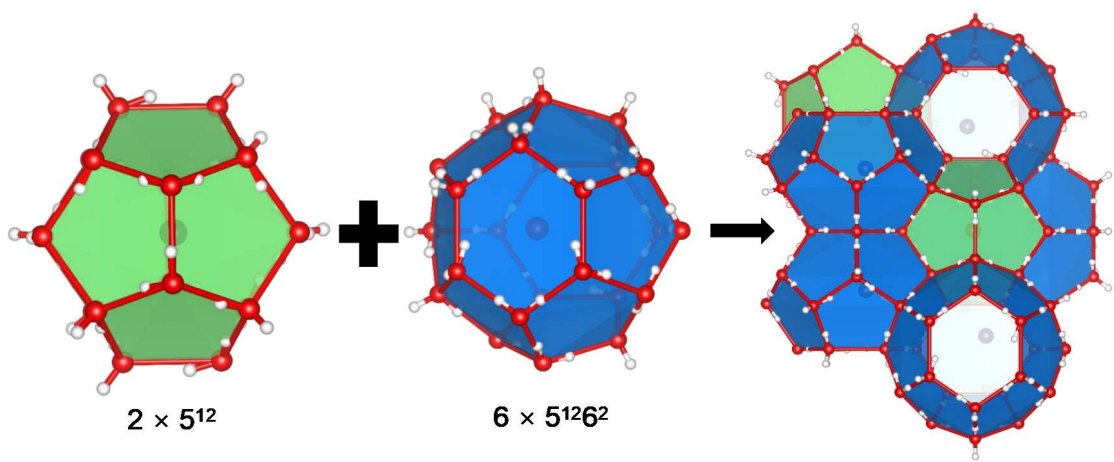

$2 \times 5^{12}$      $6 \times 5^{12}6^2$

Figure 1 Molecular structure of sI clathrate hydrate. One unit cell of sI clathrate hydrate consists of six large polyhedral cages ($5^{12}6^2$) and two small polyhedral cages ($5^{12}$) that are formed by 46 water molecules.

**2.2 Forcefield of Clathrate Hydrates**

Reasonable molecular forcefields are critical to the accuracy of MD simulations. In this work, water molecules in CHs are described by a full atomic rigid TIP4P/ICE water model[40, 41]. For the guest molecules in CHs, they are mimicked by the full atomic OPLS-AA forcefield[42]. This OPLS-AA forcefield has been proved to be highly successful in computing conformational energetics, thermodynamic and structural properties of organic liquids[11]. The intermolecular non-bonding dispersion/repulsion forces in the CH systems are described by the standard 12-6 Lennard-Jones (LJ) potential. The Lorentz-Berthelot combining rule is utilized to determine the interaction parameters between different LJ pairs. Table 2 lists the corresponding LJ parameters and atomic charges of all investigated guest molecules. The cutoff distance of the LJ interactions is set to be 12.0 Å. The particle-particle-grid (PPPM) method is applied to calculate electrostatic long-range interaction forces.

Table 2 Lennard-Jones parameters and atomic charges of OPLS-AA forcefield of all the investigated guest molecules[42]

| Molecules | Atom Type | $\varepsilon$ (kcal/mole) | $\sigma$ (Å) | $q$ (e) |
|---|---|---|---|---|
| $CH_4$ | C | 0.066 | 3.50 | -0.240 |
|  | H | 0.030 | 2.50 | 0.060 |
| $NH_3$ | N | 0.170 | 3.42 | -1.020 |
|  | H | 0.000 | 0.00 | 0.340 |
| $H_2S$ | S | 1.046 | 3.70 | -0.470 |
|  | H | 0.000 | 0.00 | 0.235 |
| $CH_2O$ | C | 0.105 | 3.75 | 0.450 |
|  | O | 0.210 | 2.96 | -0.450 |
|  | H | 0.015 | 2.42 | 0.000 |
| $CH_3OH$ | C | 0.066 | 3.50 | 0.145 |
|  | O | 0.170 | 3.12 | -0.683 |
|  | H(C) | 0.030 | 2.50 | 0.040 |
|  | H(O) | 0.000 | 0.00 | 0.418 |
| $CH_3SH$ | C | 0.066 | 3.50 | 0.000 |
|  | S | 0.250 | 3.55 | -0.435 |
|  | H(C) | 0.030 | 2.50 | 0.060 |
|  | H(S) | 0.000 | 0.00 | 0.255 |

**2.3 Tensile MD simulations**

Prior to mechanical tension, molecular configuration of CHs is quasi-statically optimized to a local minimum configuration, with energy and force tolerances of $1.0 \times 10^{-4}$ kcal/mole and $1.0 \times 10^{-4}$ kcal/(mole·Å), respectively. Then, within MD simulation time of 100 ps, as-minimized samples are heated from 1-150 K at a constant confining pressure of 0.1 MPa. Afterwards, MD calculations with simulation time of 400 ps are performed to fully relax the structures at constant temperature of 150 K and confining pressure of 0.1 MPa under NPT ensemble (constant number of particles, constant pressure, and constant temperature). The Nosé-Hoover thermostat and Nosé-Hoover barostat with damping times of 0.1 and 1.0 ps are employed to control the temperature and pressure of CH systems, respectively. Finally, deformation control method is employed to achieve uniaxial tension. The set-up of the deformation control method corresponds to a modified NPT ensemble, specifically, NVT in the loading direction, and NPT in the lateral directions, which is able to control pressure only in non-loading directions ($x$ and $y$ directions) independently and temperature. A reasonable constant strain rate of $10^8$/s is applied and the coordinates of all molecular centroids along the straining direction are readjusted uniformly every 1000 timesteps. The transverse pressures are independently maintained, ensuring that the straining samples are able to undergo expansion/contraction in the lateral directions because of Poisson effect. A timestep of 1.0 fs with Velocity-verlet algorithm is utilized to integrate the Newton's equations. All the MD calculations are performed using a large-scale atom-molecular massively parallel simulator (LAMMPS) software package code[43].

## 3 Results and Discussion

### 3.1 Effects of Guest Molecules on the Mechanical Properties of Clathrate Hydrates

Data of available mechanical properties of CHs are scarce. Figure 2 shows the simulated stress-strain curves of a variety of guest molecules-contained and guest-free sI CHs subjected to uniaxial tensile loads. Obviously, all CHs exhibit unique mechanical responses that greatly vary with the type of guest molecules, and guest-free CH exhibits similar tensile responses to those of $CH_4$, $CH_2O$, and $H_2S$ CHs. In terms of the characteristic stress-strain curves, two distinct types of mechanical responses of guest molecules-contained CHs can be roughly classified. One type is mainly represented by $CH_4$, $CH_2O$, and $H_2S$ CHs. Based on the global loading curves, four deformational stages can be roughly identified. The first deformation stage is characterized by linear increase in stress with finite increase of strain, signifying linear elastic behaviors. The deformational stage II is described by that, with increasing strain, the loading stresses nonlinearly increase up to the highest peaks, suggesting that CHs are nonlinear-elastically deformed. Based on the curves, strain-softening behaviors can be also identified. The deformational stage III is primarily characterized by sudden deep drops of loading stresses in the stress-strain curves. Such drops of loading stresses indicate occurrence of plastic deformation-induced structural destabilization of CHs. The final deformational stage is characterized by a large number of sudden rise-and-drop of tensile stress events in the following long-range curves, indicating significant plasticity. The other type is represented by $CH_3OH$, $CH_3SH$, and $NH_3$ CHs. In contrast, three deformational stages are identified from the loading curves. The deformational stage I is described by the initial short-range linear stress-strain relations, implying linear elastic responses. In deformational stage II, the tensile stresses nonlinearly increase up to maximum values within finite strains. Remarkably, the nonlinearity in those tensile stress-strain curves is characterized by a series of rise-and-drop in stress events, differing from the cases of $CH_4$, $CH_2O$, and $H_2S$ CHs, which is characteristic plastic deformations. Similarly, the final deformational stage is characterized by large-scale rise-and-drop of loading stress events in the curves. It is summarized that both types of CHs exhibit distinct plastic deformation mechanisms.

To reveal the influence of temperature on their tensile properties, the tensile stress-strain curves of all six guest molecules-contained CHs at high temperatures of 200 K and 250 K are examined as shown in Figures S1 and S2 of Supporting Information. Overall, the characteristics of tensile loading responses of stable $CH_4$, $CH_2O$, $H_2S$ and $CH_3SH$ CHs are negligibly influenced by the external temperature. With regard to CHs containing $NH_3$ and $CH_3OH$ guest molecules that have strong ability to form hydrogen bonds with water molecule and strong molecular polarity; however, they show zero-tensile stress during the entire loading process at high external temperatures. This indicates that both $NH_3$ and $CH_3OH$ CHs are not structural stable at high temperatures.

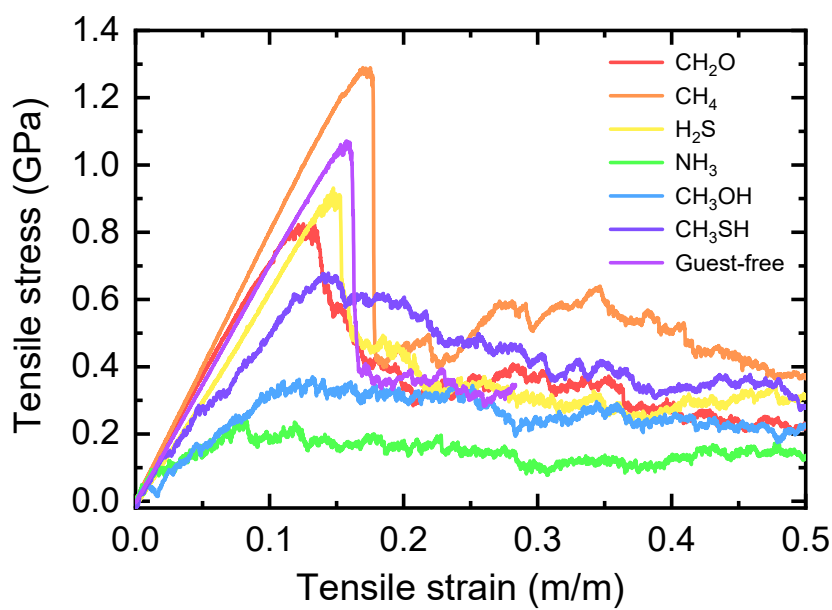

Figure 2 Mechanical stress-strain curves of clathrate hydrates encapsulating six different guest molecules of $CH_2O$, $CH_4$, $H_2S$, $NH_3$, $CH_3OH$ and $CH_3SH$, respectively, as well as that of guest-free clathrate hydrate for comparison.

Table 3 lists the Young's modulus and tensile limit of those six guest molecules-contained CHs obtained from the stress-strain curves of Figure 2, as well as guest-free CH. In terms of the values of Young's modulus, they are sorted as $CH_4$ > $CH_3SH$ > $CH_2O$ > Guest-free > $NH_3$ > $H_2S$ > $CH_3OH$ CHs, with maximum and minimum values of 8.26 GPa and 5.50 GPa, respectively. In terms of the values of tensile limit, they are ranked as $CH_4$ > Guest-free > $H_2S$ > $CH_2O$ > $CH_3SH$ > $CH_3OH$ > $NH_3$ CHs, which is different from the case of Young's modulus. In short, $CH_4$ CH is the most mechanically robust clathrate structure. This demonstrate the fact that sI $CH_4$ CH predominates in natural settings. Overall, the guest molecular size, geometric configuration, molecular polarity, as well as the ability of guest molecules to form H-bonds are primarily attributed to the differences in their tensile properties.

Table 3 Young's moduli and tensile limits of clathrate hydrates encapsulating six different guest molecules, as well as guest-free clathrate hydrate.

| Tensile properties | $CH_2O$ | $CH_4$ | $H_2S$ | $NH_3$ | $CH_3OH$ | $CH_3SH$ | Guest-free |
|---|---|---|---|---|---|---|---|
| Young's modulus (GPa) | 7.60 | 8.26 | 6.09 | 6.56 | 5.50 | 7.83 | 6.88 |
| Tensile limit (GPa) | 0.82 | 1.20 | 0.92 | 0.24 | 0.36 | 0.67 | 1.07 |

**3.2 Radial Distribution Functions (RDFs) and Hydrogen-bonding Configurations in Clathrate Hydrates**

The different mechanical properties of the six CHs can be indicative of their intrinsic molecular structures. To characterize their molecular structures, RDFs of relaxed CHs are computed. Figure 3a shows the RDFs of oxygen-oxygen ($O_W\cdots O_W$) of host water molecules in the CHs containing six different guest molecules, as well as the case of guest-free CH for comparison. Apparently, the RDFs curves show a number of similar peaks in the distance of 0-10 Å, demonstrating the characteristics of crystalline structures. This indicates that all investigated guest molecules contained and guest-free CHs preserve sI clathrate frameworks formed by water molecules. By comparison, there are slight differences in the positions and values of peaks between RDFs of $O_W\cdots O_W$ of CHs. For example, the values of the second, third and fourth peaks are varied with guest molecular types, all with minimum and maximum values for $NH_3$ and $CH_4$ CHs, respectively. Moreover, there is a slight shift in the positions of peaks between those RDFs. This indicates that there exists difference in the clathrate structures between CHs containing different guest molecules, resulting in their different mechanical properties. Figure 3b displays the RDFs of host water and guest molecules ($O_W \cdots G_W$) for CHs containing six different guest molecules. Similarly, except for $NH_3$ CH, there are a number of peaks identified in the RDFs curves, indicating that they are uniformly distributed in the clathrate cages of CHs. The position values of the first peak in the RDFs of $O_W \cdots G_W$ represent mean distances between guest and host molecules. The asymmetry of the peaks in the RDFs mainly comes from the fact that guest molecules occupy different $5^{12}6^2$ and $5^{12}$ clathrate cages[44]. Interestingly, small guest $NH_3$ CH shows distinct RDFs of $O_W \cdots G_W$ from those of other CHs. For example, within distance of 2-5 Å, there are several weak peaks identified in the RDF curve, suggesting that $NH_3$ guest molecules do not prefer staying in the center of clathrate cages, but prefer approaching local host molecules of clathrate cages for more easily forming H-bonds with surrounding water molecules, resulting in its weak mechanical properties. As a result of strong ability of $NH_3$ to form hydrogen bonds with water molecule, $NH_3$ has been assigned a main role as water-ice antifreeze and methane hydrate inhibitor[45]. However, $NH_3$ is a small molecule that is suitably enclathrated by $5^{12}$ and $5^{12}6^2$ polyhedral cages. Thus, from the dimensionality alone, it possesses the great potential to be a suitable guest molecule for cages of sI CH[46]. Those are mainly responsible for the distinct tensile properties and RDFs of $NH_3$ CH from those of other five CHs. The specific behaviors of $NH_3$ CH such as the RDFs can be utilized for identification of $NH_3$ CH from a variety of CHs, as well as playing an understood role in the extraterrestrial space. With regard to CHs containing $CH_3OH$, $CH_3SH$ guest molecules that also have ability to form H-bonds with water molecules; however, as a result of their unique molecular configurations, $CH_3OH$, $CH_3SH$ guest molecules are not able to approach local host water molecules of finite-dimensional clathrate cages, leading to single peak in the distance of 2-5 Å. This explains that the guest molecular polarity and geometrical configuration dictate the mechanical properties of sI CHs.

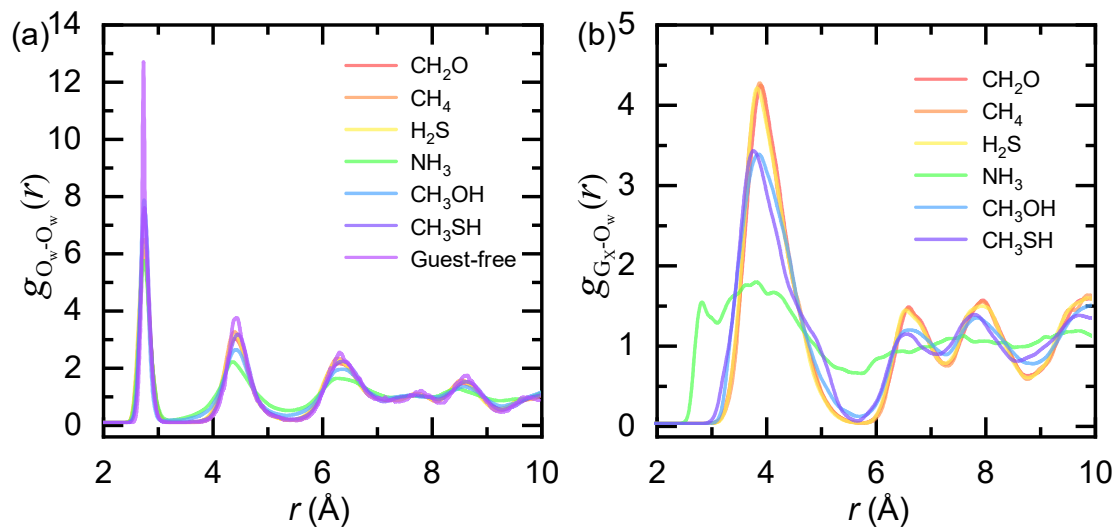

Figure 3 RDFs of relaxed clathrate hydrates containing $CH_2O$, $CH_4$, $H_2S$, $NH_3$, $CH_3OH$, $CH_3SH$ guest molecules, as well as guest-free clathrate structure. (a) RDFs of oxygen of water molecules ($O_W\cdots O_W$) of clathrate hydrates and (b) RDFs of guest molecule-oxygen of water molecules ($G_X\cdots O_W$) in relaxed clathrate hydrates entrapping six different guest molecules.

To further reveal the structural characteristics of those six guest molecules-contained CHs, the motifs of $5^{12}$ and $5^{12}6^2$ polyhedral cages encapsulating guest molecules at equilibrium state are captured in Figure 4. Apparently, the captured water frameworks preserve the motifs of $5^{12}$ and $5^{12}6^2$ clathrate cages, indicating that all investigated CHs are structurally stable structures. However, the geometrical configuration of $5^{12}$ and $5^{12}6^2$ water cages varies with the encapsulated guest molecules. For example, small $5^{12}$ clathrate cage entrapping small $NH_3$ molecule that has strong ability of forming hydrogen bonds with water molecule and strong molecular polarity is more pronounced distorted than that entrapping other guest molecules. Whereas, large $5^{12}62$ clathrate cage encapsulating either small $NH_3$ or large $CH_3OH$ guest molecules that show strong H-bonding formation ability with water molecule and strong molecular polarity is also more significant distorted.

This indicates that guest molecules with strong ability of forming H-bonds with water molecules and strong molecular polarity can result in significant localized distortion of clathrate cages. Except for $CH_4$ guest molecule, other five guest molecules are able to form H-bonds with water molecules of both $5^{12}$ and $5^{12}6^2$ clathrate cages, while their H-bonding configurations are distinct from each other.

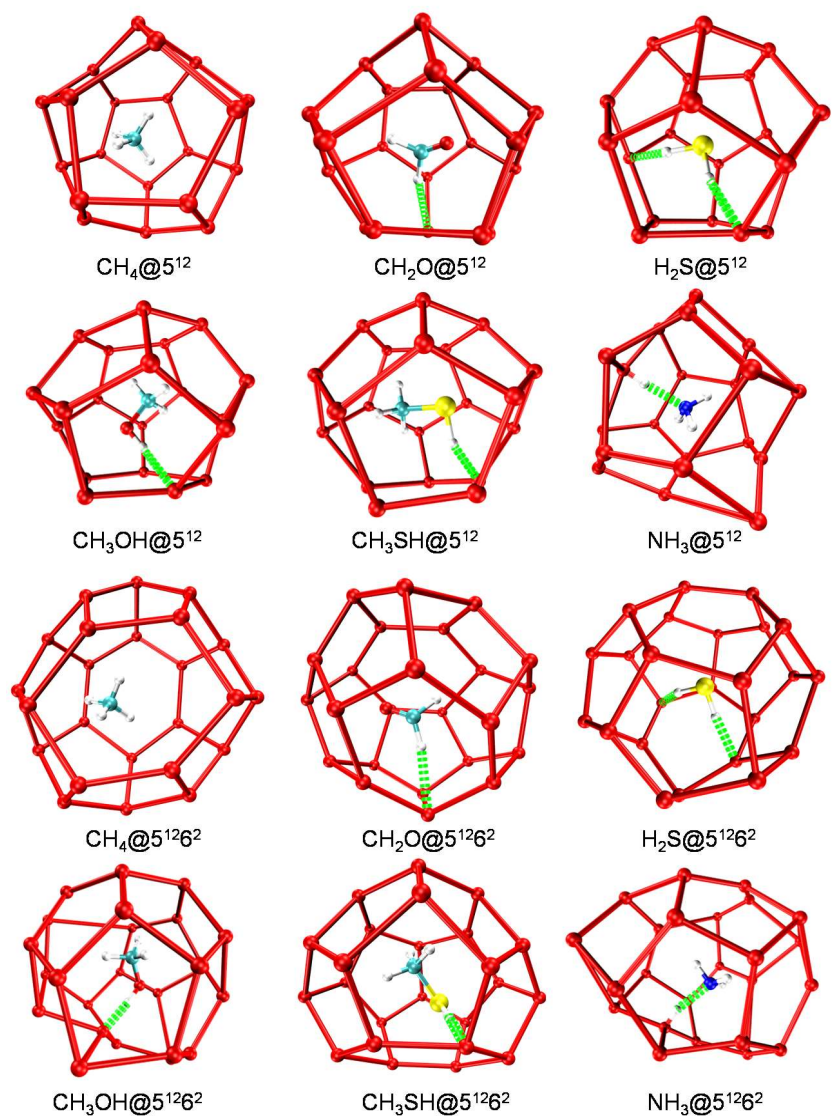

Figure 4 Perspective motifs of guest molecules $CH_4$, $CH_2O$, $H_2S$, $CH_3OH$, $CH_3SH$ and $NH_3@5^{12}/5^{12}6^2$ of relaxed clathrate hydrates. To clarify the clathrate cages, hydrogen atoms of water molecules are removed and neighboring oxygen of water molecules are connected. The hydrogen bonds of host water-guest molecules are highlighted by green-springs.

**3.3 Energetics in Deformed Clathrate Hydrates**

Non-bonded intermolecular interactions play key roles in the stabilizing clathrate cages and mechanical stability of CHs. Therefore, the total potential energies of nonbonded intermolecular interactions of host-guest ($E_{\text{host-guest}}$) in deformed CHs are computed. Figure 5 shows variations in $E_{\text{host-guest}}$ of six different CHs with strain. Apparently, all CHs show different $E_{\text{host-guest}}$ during the global deformation and uniquely nonlinear $E_{\text{host-guest}}$-strain curves. In terms of $E_{\text{host-guest}}$ at zero strain, they are sorted as $NH_3$ > $CH_3OH$ > $CH_3SH$ > $H_2S$ > $CH_2O$ > $CH_4$. Such sorting is almost opposite to the sorting in terms of mechanical properties. This indicates that strong nonbonded interactions of host-guest molecules in CHs do not enhance structural stability but degrade the mechanical performance. As seen in Figure 5, $E_{\text{host-guest}}$ vary with strain. As for $CH_4$ CH, it is observed that $E_{\text{host-guest}}$ nonlinearly increases with increasing strain, indicating that elongation of CHs globally reduces the intermolecular interactions between $H_2O$ and $CH_4$ molecules. This is mainly due to the fact that the $CH_4$ molecule is a non-polar molecule and does not have ability to form H-bonds with water molecules. As the $CH_4$ CH is elastically strained, intermolecular interactions between $H_2O$ and $CH_4$ molecules become less and less pronounced. The first drop of $E_{\text{host-guest}}$ in the curve occurs at around strain of 0.16 that is the failure strain. As the $CH_4$ CH is plastically deformed, intermolecular interactions between $H_2O$ and $CH_4$ molecules are further weakened. With regard to other CHs, however, clear reduction tendency in the $E_{\text{host-guest}}$ with strain, representing that deformation of CHs enhances host-guest intermolecular interactions, which is in contrast to the case of $CH_4$ CH. As for CHs containing polar $CH_3SH$, $CH_3OH$ and $NH_3$ guest molecules that have strong ability to form H-bonds with water molecules, it is observed that $E_{\text{host-guest}}$ sharply reduces in the early elongation, but then gradually approaches to constant values with increasing strain. This indicates that, prior to failure, deformation of CHs significantly enhances intermolecular interactions of host-guest molecules, thereby destabilizing the clathrate cages formed by water molecules. With regard to CHs containing polar $CH_2O$ and $H_2S$ guest molecules that have weak ability to form H-bonds with water molecules, it is detected that, prior to failure, $E_{host-guest}$ relatively insignificantly varies with elastic strain. Intriguingly, there is a crossover in $E_{host-guest}$ at a critical elastic strain for $H_2S$ CH. As $H_2S$ CH is elastically strained, $E_{host-guest}$ first slightly increases but then decreases, differing from $CH_2O$ CH that show monotonic reduction in $E_{host-guest}$. However, both $CH_2O$ and $H_2S$ CHs show sudden deep drop of $E_{host-guest}$ at critical strains that correspond to the failure strains. With increasing strain, $E_{host-guest}$ of $CH_2O$ and $H_2S$ CHs further decreases.

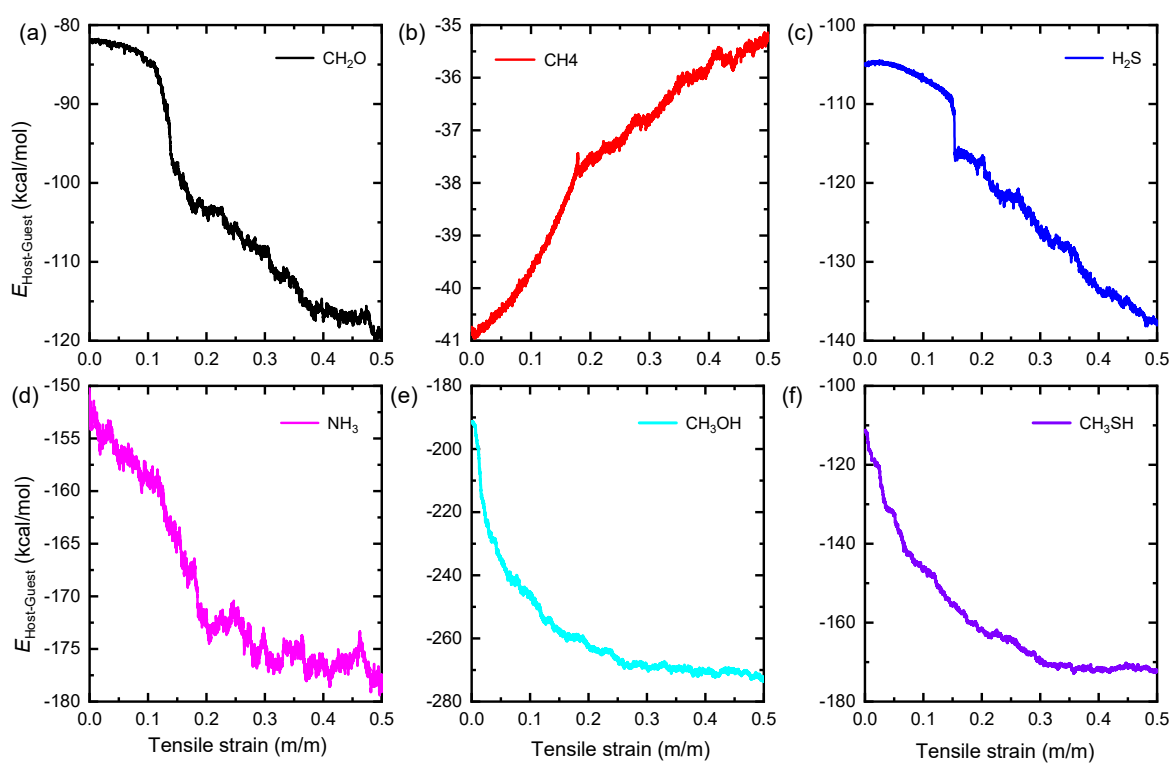

Figure 5 Energetics of non-bonded interactions of host-guest molecules in clathrate hydrates. Variations in total potential energies of host water-guest molecules of (a) $CH_2O$, (b) $CH_4$, (c) $H_2S$, (d) $NH_3$, (e) $CH_3OH$ and (f) $CH_3SH$ in clathrate hydrates with strain, respectively.

## 3.4 H-bonds in Deformed Clathrate Hydrates

In CHs, host water molecules are uniquely connected by H-bonds to form the basic skeleton of hydrate structures[47]. H-bonding network is the key to stabilize CH structures. Quantitative analysis of H-bonds in CHs helps to understand the stability and mechanical performance of CHs subjected mechanical loads. Figure 6a shows schematic diagram of identification of H-bonds for water systems. To determine a H-bond, cutoff values of the donor-acceptor angle $\alpha$ and interatomic distance $R$ are set to be 30° and 3.5 Å, respectively. Figure 6b shows the development of total number of H-bonds identified in CHs containing $NH_3$, $CH_3OH$, $CH_3SH$, $H_2S$, $CH_4$, and $CH_2O$ guest molecules during the whole straining process. Obviously, the polarity and dimensionality of guest molecules in CHs significantly influences the total number of H-bonds, although CHs are composed of identical number of water molecules that form H-bonded framework. At zero strain, in terms of number of H-bonds, they are sorted as $NH_3$ > $CH_3OH$ > $CH_3SH$ > $H_2S$ > $CH_4$ > $CH_2O$ CHs. This clearly suggests that, except for $CH_4$ and $CH_2O$ CHs, there are a number of H-bonds formed between host water and guest molecules, which closely correlates with the dimensionality, polarity of guest molecules, and the ability to form H-bonds. As the CHs are strained, there is rise-and-drop in the total number of H-bonds, indicating that H-bonds in CHs are able to dynamically dissociate and form with deformation. Interestingly, there is a rise tendency in the total number of H-bonds with increasing strain for CHs containing small guest molecules, whereas for other CHs, the total number of H-bonds tends to decrease. Figure 6c shows variations in the number of H-bonds formed between host water molecules in CHs containing $CH_2O$, $CH_4$, $H_2S$, $NH_3$, $CH_3OH$, $CH_3SH$ guest molecules. Apparently, identical host water molecules of the six different CHs form different number of H-bonds, indicating that H-bonds formed by water molecules in CHs are greatly affected by the properties of guest molecules such as molecular dimensionality, atomic charges, polarity, the ability to form H-bonds with water molecules, and so on. It is observed that, during the entire stretching process, there is reduction trend in the number of H-bonds formed between host water molecules, suggesting that, beside the external conditions including temperature and confining pressure, mechanical deformation also cause dissociation of H-bonded network of CHs. H-bonds formed between water molecules in CHs containing large $CH_3OH$, $CH_3SH$ guest molecules are more sensitive to mechanical deformation, whereas for $CH_4$ CH, they are the least sensitive to strain, explaining that $CH_4$ CH is the most mechanically robust structure. Figure 6d displays the variation in the number of H-bonds formed between host water and guest molecules in CHs containing $CH_2O$, $CH_4$, $H_2S$, $NH_3$, $CH_3OH$, $CH_3SH$ guest molecules with strain. Apparently, those guest molecules encapsulating in water cages of sI CH present distinct abilities in forming H-bonds with host water molecules. At equilibrium state, in terms of guest molecular ability to form H-bonds with water molecules, they are ranked as $NH_3 > CH_3OH > CH_3SH > H_2S > CH_2O = CH_4$. Such ranking clearly reveals that sI CHs containing guest molecules with stronger ability in forming H-bonds with water molecules of clathrate cages show poorer mechanical properties. This is because, once guest and host molecules in CHs form H-bonds, H-bonds formed between host water molecules in CHs are limited, resulting in reduction in the number of H-bonds in the basic H-bonded skeleton structure, thereby weakening the mechanical performance. Another reason is that guest molecules with strong ability in forming H-bonds tend to approach the edge of water cages instead of stay in the center of water cages, generating localized distortion of water cages, thereby destabilizing the water cages. Except for $CH_4$ CH, as the samples are strained, the number of H-bonds of host-guest molecules in investigated CHs greatly varies. It is also observed that there is global increase in the number of

H-bonds of host-guest molecules during the whole straining process. This is due to the fact that deformation of clathrate cages reduces the distance of potential H-bonding donor-acceptor, thereby enhancing the formation of host-guest molecular H-bonds. As the CHs containing $H_2S$ and $CH_2O$ guest molecules are elastically deformed, there is negligible change in the number of H-bonds of host-guest molecules; however, as the CHs are plastically stretched, there is rising tendency in the number of H-bonds of host-guest molecules. This comes from the fact that, upon elastic straining, the distance of guest $H_2S$ and $CH_2O$ molecules to host water molecules is over cutoff value (3.5 Å) of H-bonding donor-acceptor, whereas upon plastic deformation, host water molecules from dissociated clathrate cages are free to approach guest $H_2S$ and $CH_2O$ molecules. Note that, because CH4 molecule does not have donor and acceptor of H-bond, there is null in the number of H-bonds of host-guest molecules during the whole stretching process.

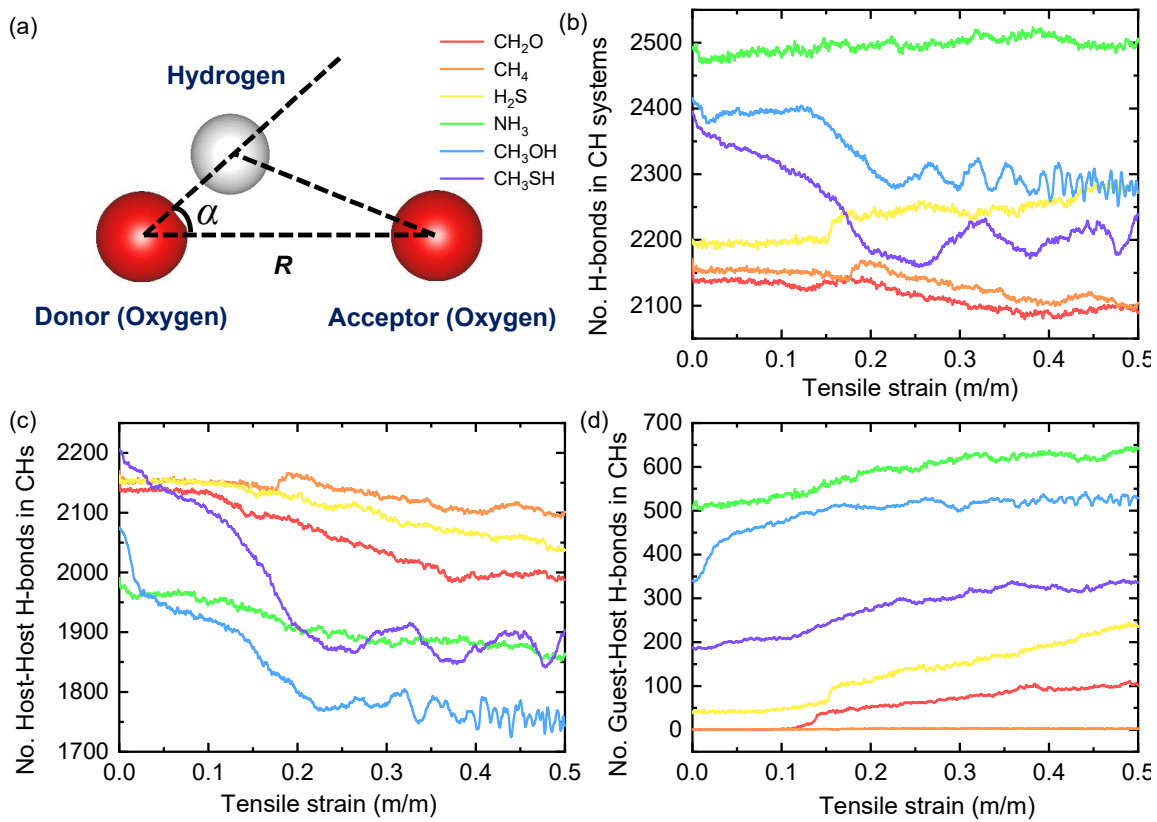

Figure 6 Hydrogen bonds in deformed clathrate hydrates. (a) Schematic diagram of determination

criteria of hydrogen bonds. Variations in the (b) total number of hydrogen bonds, (c) number of host-host hydrogen bonds and (d) number of guest-host hydrogen bonds in clathrate hydrates entrapping $NH_3$, $CH_3OH$, $CH_3SH$, $H_2S$, $CH_4$, and $CH_2O$ guest molecules with strain, respectively.

**3.5 Strain-induced Instability in Clathrate Hydrate**

To understand the instability of all investigated CHs subjected mechanical load, the development of molecular structures during the whole stretching process are recorded. Figure 7 shows a series of side-viewed snapshots of the investigated CHs where the oxygen atoms of host water molecules are rendered according to their values of shear strain. Clearly, there are two distinct mechanical destabilization patterns, depending on the polarity and configuration of guest molecules. One is represented by $CH_4$, $CH_2O$, and $H_2S$ CHs in which the guest molecules are less molecular polarity and possess weak H-bonding ability with host water molecules. As is shown in Figure 6a-c, the mechanical destabilization occurs via brittle fracture of clathrate cages on the (101) crystalline plane, resulting in sudden deep drops of loading stress in the curves of Figure 2. The brittle fracture is primarily dominated by the local dissociation of H-bonds of clathrate cages. The other one is represented by $CH_3SH$, $CH_3OH$, and $NH_3$ CHs in which guest molecules are stronger molecular polarity, and show strong H-bonding ability. In contrast, the mechanical destabilization easily takes place through ductile-like failure of clathrate cages. Such ductile-like failure is characterized by gradual global amorphization as a result of large-scale dissociation of H-bonds in clathrate cages. This explains their unique nonlinear mechanical responses as shown in Figure 2. It is summarized that those two distinct mechanical destabilizations closely correlate with the H-bonding ability of host-guest molecules and polarity of guest molecules.

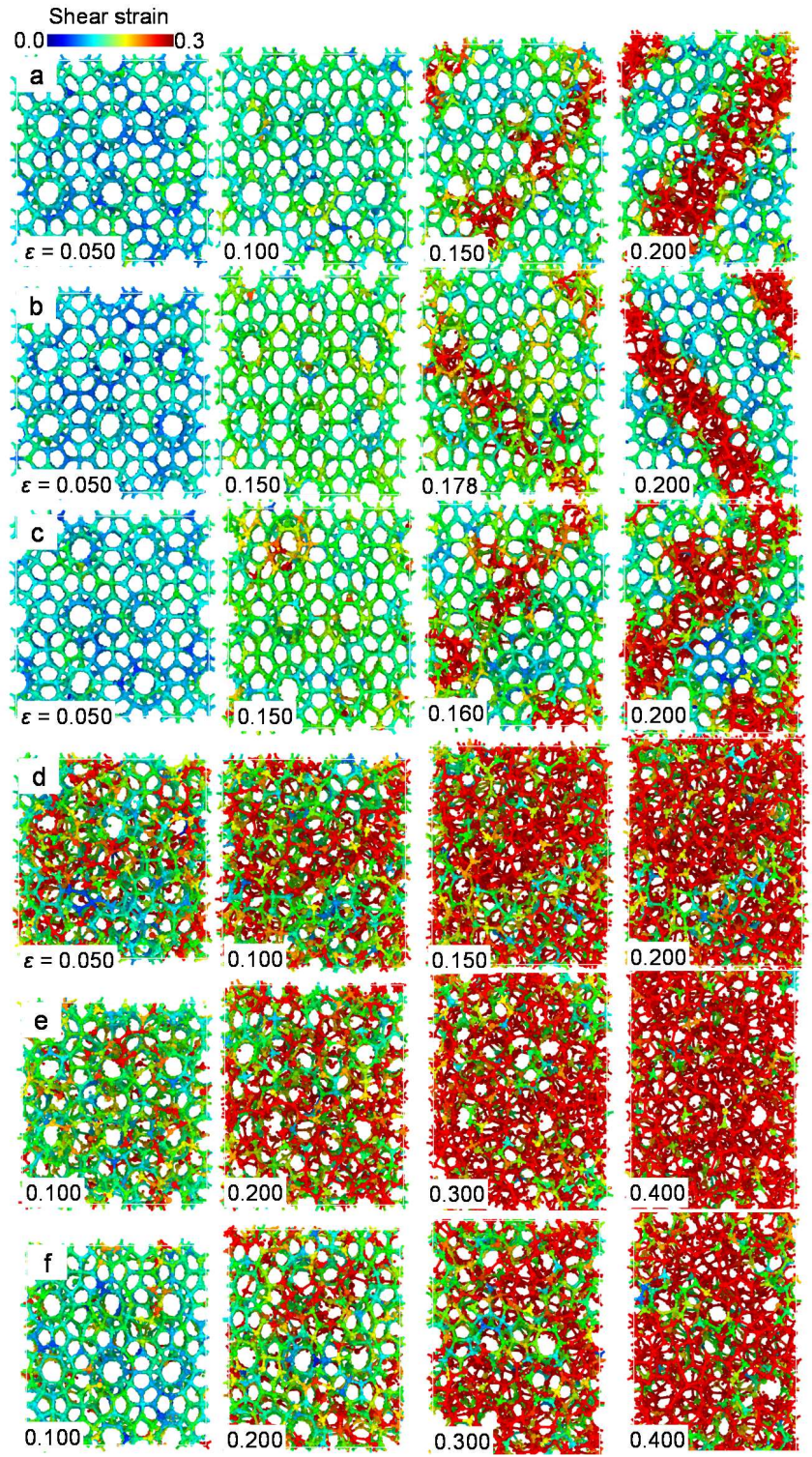

Figure 7 Uniaxial tension-induced failure in sI clathrate hydrates. Side views of representative snapshots of (a) $CH_2O$, (b) $CH_4$, (c) $H_2S$, (d) $NH_3$, (e) $CH_3OH$, and (f) $CH_3SH$ clathrate hydrates during the deformation process, where the guest molecules and the hydrogen of host water molecules are removed for clarification. Note that the uniaxial straining is along the vertical direction. The color

code is on the basis of the values of shear strain.

## 4. Conclusions

In summary, classical MD simulations are performed to examine the mechanical characteristics of sI CHs entrapping a variety of guest molecules ($CH_4$, $NH_3$, $H_2S$, $CH_2O$, $CH_3OH$, and $CH_3SH$) that show different H-bonding abilities with host water molecules subjected to uniaxial tensile loads. RDFs analysis demonstrates that all studied CHs are structurally stable host-host H-bonded network of clathrate structures, although there are H-bonds formed between host-guest molecules. Tension MD simulations show that all CHs present unique mechanical responses that greatly vary with molecular configuration, molecular polarity and forming H-bonding ability with water molecule of guest species. It is revealed that the mechanical properties of CHs such as tensile limit, failure strain, Young's modulus and destabilization pattern are dominated by the guest molecular property such as the H-bonding ability with water, molecular polarity and so on. For example, CHs entrapping $CH_4$, $H_2S$ and $CH_2O$ guest molecules that have weak ability to form H-bonds with water and weak molecular polarity show superior mechanical properties over those containing $NH_3$, $CH_3OH$ and $CH_3SH$ guest molecules that show strong H-bonding ability with water and strong molecular polarity. This mainly comes from the fact that formation of H-bonds of host-guest molecules limits formation H-bonds between host water molecules, thereby reducing the number of H-bonds in the basic H-bonded clathrate structure. Interestingly, energetics of non-bonded host-guest interactions in mechanically robust $CH_4$ CH are different from those of other CHs. Upon critical strains, CHs entrapping $CH_4$, $H_2S$ and $CH_2O$ guest molecules fail via brittle failure on the (101) crystalline plane, whereas for other CHs, they are mechanically destabilized via ductile failure due to gradual global dissociation of H-bonds in clathrate cages. Moreover, the deformation of CHs enhances the host-guest H-bonds. This study provides important insights into the mechanical stability and deformation mechanisms of clathrate hydrate structures.

## Acknowledgments

This work is financially supported by the National Natural Science Foundation of China (Grant Nos. 11772278, 11904300 and 11502221), the Jiangxi Provincial Outstanding Young Talents Program (Grant No. 20192BCBL23029), the Fundamental Research Funds for the Central Universities (Xiamen University: Grant Nos. 20720180014, 20720180018 and 20720180066). Y. Yu and Z. Xu from Information and Network Center of Xiamen University for the help with the high-performance computer.

## References

1. T. S. Collett, Aapg Bull **86** (11), 1971-1992 (2002).
2. E. D. Sloan, Nature **426** (6964), 353-359 (2003).
3. K. A. Kvenvolden, Rev Geophys **31** (2), 173-187 (1993).
4. G. R. Dickens, C. K. Paull and P. Wallace, Nature **385** (6615), 426-428 (1997).
5. W. Y. Xu, R. P. Lowell and E. T. Peltzer, Journal Of Geophysical Research-Solid Earth **106** (B11), 26413-26423 (2001).
6. K. A. Kvenvolden, International Conference on Natural Gas Hydrates **715**, 232-246 (1994).
7. L. E. Zerpa, J. L. Salager, C. A. Koh, E. D. Sloan and A. K. Sum, Industrial & Engineering Chemistry Research **50** (1), 188-197 (2011).
8. A. Kumar, O. S. Kushwaha, P. Rangsunvigit, P. Linga and R. Kumar, Can J Chem Eng **94** (11), 2160-2167 (2016).
9. H. Mimachi, S. Takeya, A. Yoneyama, K. Hyodo, T. Takeda, Y. Gotoh and T. Murayama, Chem Eng Sci **118**, 208-213 (2014).
10. A. Kumar, H. P. Veluswamy, R. Kumar and P. Linga, Applied Energy **235**, 21-30 (2019).
11. H. Komatsu, K. Maruyama, K. Yamagiwa and H. Tajima, Chemical Engineering Research and Design **150**, 289-298 (2019).
12. D. W. Davidson, Y. P. Handa, C. I. Ratcliffe, J. S. Tse and B. M. Powell, Nature **311** (5982), 142-143 (1984).
13. R. L. Christiansen and E. D. Sloan, International Conference on Natural Gas Hydrates **715**, 283-305 (1994).
14. M. Arjmandi, A. Chapoy and B. Tohidi, Journal Of Chemical And Engineering Data **52** (6), 2153-2158 (2007).
15. B. Kvamme and O. K. Forrisdahl, Fluid Phase Equilibr **83**, 427-435 (1993).
16. B. Kvamme, A. Lund and T. Hertzberg, Fluid Phase Equilibr **90** (1), 15-44 (1993).
17. D. W. Davidson, Y. P. Handa, C. I. Ratcliffe, J. A. Ripmeester, J. S. Tse, J. R. Dahn, F. Lee and L. D. Calvert, Mol Cryst Liq Cryst **141** (1-2), 141-149 (1986).
18. H. Tanaka, Y. Tamai and K. Koga, J Phys Chem B **101** (33), 6560-6565 (1997).
19. J. A. Ripmeester and C. I. Ratcliffe, J Phys Chem-Us **94** (25), 8773-8776 (1990).
20. J. X. Liu, Y. J. Yan, J. F. Xu, S. J. Li, G. Chen and J. Zhang, Computational Materials Science **123**, 106-110 (2016).

21. R. Susilo, S. Alavi, I. L. Moudrakovski, P. Englezos and J. A. Ripmeester, Chemphyschem **10** (5), 824-829 (2009).
22. S. Alavi, K. Shin and J. A. Ripmeester, Journal Of Chemical And Engineering Data **60** (2), 389-397 (2015).
23. L. A. Stern, S. H. Kirby and W. B. Durham, Science **273** (5283), 1843-1848 (1996).
24. L. A. Stern, S. H. Kirby and W. B. Durham, Energy & Fuels **12** (2), 201-211 (1998).
25. T. M. Vlasic, P. D. Servio and A. D. Rey, Crystal Growth & Design **17** (12), 6407-6416 (2017).
26. J. H. Jia, Y. F. Liang, T. Tsuji, S. Murata and T. Matsuoka, Scientific Reports **7** (2017).
27. Q. Shi, P. Q. Cao, Z. D. Han, F. L. Ning, H. Gong, Y. Xin, Z. S. Zhang and J. Y. Wu, Cryst Growth Des **18** (11), 6729-6741 (2018).
28. R. K. Mcmullan and G. A. Jeffrey, J Chem Phys **42** (8), 2725-2732 (1965).
29. F. Takeuchi, M. Hiratsuka, R. Ohmura, S. Alavi, A. K. Sum and K. Yasuoka, Journal Of Chemical Physics **138** (12) (2013).
30. J. J. Shieh and T. S. Chung, Journal of Polymer Science Part B: Polymer Physics **37** (20), 2851-2861 (1999).
31. W.-H. Lin and T.-S. Chung, Journal of Membrane Science **186** (2), 183-193 (2001).
32. T. Yoshioka, M. Kanezashi and T. Tsuru, AIChE Journal **59** (6), 2179-2194 (2013).
33. T. Zhou, Y. Sang, X. Wang, C. Wu, D. Zeng and C. Xie, Sensors and Actuators B: Chemical **258**, 1099-1106 (2018).
34. J.-R. Li, R. J. Kuppler and H.-C. Zhou, Chemical Society Reviews **38** (5), 1477-1504 (2009).
35. S. Goel, Z. Wu, S. I. Zones and E. Iglesia, Journal of the American Chemical Society **134** (42), 17688-17695 (2012).
36. C. Sun, B. Wen and B. Bai, Chemical Engineering Science **138**, 616-621 (2015).
37. A. Koriakin, Y.-H. Kim and C.-H. Lee, Industrial & engineering chemistry research **51** (44), 14489-14495 (2012).
38. P. Maksymovych, D. C. Sorescu, D. Dougherty and J. T. Yates, The Journal of Physical Chemistry B **109** (47), 22463-22468 (2005).
39. A. T. Güntner, S. Abegg, K. Wegner and S. E. Pratsinis, Sensors and Actuators B: Chemical **257**, 916-923 (2018).
40. L. C. Jacobson and V. Molinero, Journal Of Physical Chemistry B **114** (21), 7302-7311 (2010).
41. J. L. F. Abascal, E. Sanz, R. G. Fernandez and C. Vega, Journal Of Chemical Physics **122** (23) (2005).
42. W. L. Jorgensen, D. S. Maxwell and J. TiradoRives, Journal Of the American Chemical Society **118** (45), 11225-11236 (1996).
43. B. FrantzDale, S. J. Plimpton and M. S. Shephard, Engineering with Computers **26** (2), 205-211 (2010).
44. N. J. English and J. M. D. MacElroy, Journal Of Computational Chemistry **24** (13), 1569-1581 (2003).
45. K. Shin, R. Kumar, K. A. Udachin, S. Alavi and J. A. Ripmeester, Proceedings of the National Academy of Sciences **109** (37), 14785-14790 (2012).
46. W. L. Mao, H.-k. Mao, A. F. Goncharov, V. V. Struzhkin, Q. Guo, J. Hu, J. Shu, R. J. Hemley, M. Somayazulu and Y. Zhao, Science **297** (5590), 2247-2249 (2002).
47. G. A. Jeffrey, Advances In Enzymology And Related Areas Of Molecular Biology **65**, 217-254 (1992).